\documentclass{jpp}
\pdfoutput=1

\usepackage{amsmath}
\usepackage{epstopdf}
\usepackage[toc]{appendix}
\usepackage{color}
\usepackage{lineno}
\usepackage{bm}
\renewcommand{\thesubsection}{\Roman{subsection}}
\usepackage{graphicx}
\usepackage[utf8]{inputenc}
\usepackage[T1]{fontenc}

\shorttitle{PIC simulations of AW PDI in a low-beta plasma}
\shortauthor{C.A. Gonz\'alez, M.E. Innocenti and A. Tenerani}

\title{Particle-in-cell simulations of Alfv\'en wave parametric decay in a low-beta plasma}

\author{C.A. González \aff{1} \corresp{\email{carlos.gonzalez1@austin.utexas.edu}}, Maria Elena Innocenti\aff{2}
\and Anna Tenerani\aff{1}
}

\affiliation{
\aff{1}Department of Physics, The University of Texas at Austin, Austin, TX 78712, USA
\aff{2}Institut f{\"u}r Theoretische Physik, Ruhr-Universit{\"a}t Bochum, Bochum, Germany
}

\begin{document}

\maketitle

\begin{abstract}
We study the parametric decay instability of parallel propagating Alfvén wave in a low-beta plasma using one-dimensional fully kinetic simulations. We focus for the first time on the conversion of the energy stored in the initial Alfv\'en wave into particle internal energy, and on its partition between particle species. We show that compressible fluctuations generated by the decay of the pump wave into a secondary ion-acoustic mode and a reflected Alfv\'en wave contribute to the gain of internal energy via two distinct mechanisms. First, the ion-acoustic mode leads nonlinearly to proton trapping and proton phase space mixing, in agreement with previous work based on hybrid simulations. Second, during the nonlinear stage, a compressible front of the fast type  develops  at the steepened edge of the backward Alfv\'en wave leading to a field-aligned proton beam propagating backwards at the Alfv\'en speed.  We find that parametric decay heats preferentially protons, which gain about 50\% of the pump wave energy in the form of internal energy. However, we find that electrons are also energized and that they contribute to the total energy balance by gaining 10\% of the pump wave energy. By investigating energy partition and particle heating during parametric decay, our results contribute to determine the role of compressible and kinetic effects in wave-driven models of the solar wind. 
\end{abstract}

\section{Introduction} \label{sec:intro}
Alfvén waves are ubiquitous in magnetized plasmas and represent one of the building blocks of many phenomena in space and in laboratory. Alfv\'en waves are also thought to play a crucial role in coronal heating and solar wind acceleration. By propagating almost undisturbed along the magnetic field,  they  provide a means to transport  energy from the photosphere to the corona and further out~\citep{delzanna_2002}. Alfv\'en wave-like correlation of magnetic and velocity field fluctuations are  commonly observed in the solar wind~\citep{coleman1967,belcher1971}, providing support to wave-driven wind theories (see, e.g.,~\cite{cranmer2012self}). Understanding how the energy carried by Alfv\'en waves is ultimately converted and released to the plasma in the form of internal energy is therefore key to our understanding of coronal heating and solar wind acceleration. 

Large amplitude Alfv\'en waves at Magnetohydrodynamic scales are unstable to the Parametric Decay Instability (PDI hereafter).  In a low beta plasma (when the thermal pressure is smaller than 
magnetic pressure), PDI leads to the decay of the  ``pump'' wave  into a forward ion acoustic wave and a lower frequency backward Alfv\'en wave \citep{galeev_sov_phys_1963,Derby_1978}. The PDI is of great interest, first, because it provides a mechanism to generate reflected Alfv\'en waves which are essential to trigger the turbulent cascade~\citep{chandran_2018,malara2022parametric} -- the latter providing a possible mechanism to heat and accelerate the plasma (e.g., \cite{perez2013direct}). Second, the PDI generates also compressible modes that in turn contribute to heat the plasma via formation of shocks~\citep{del2001parametric} or, when proton kinetic effects are retained, via particle trapping and phase space mixing \citep{araneda2008proton,MatteiniEA2010}, and proton energization at steepened fronts  \citep{Gonzalez_2020,Gonzalez_2021}. Since the growth rate increases with decreasing  plasma beta, theory and simulations predict that this instability should contribute to the dynamics of the solar wind  in the inner heliosphere, preferentially in regions close to the sun where $\beta\ll1$~\citep{anna1,reville2018parametric}. Recently, signatures consistent with PDI in the lower solar atmosphere have been reported \citep{hahn_2022}, providing support to scenarios that invoke parametric decay, and in general the generation of compressible modes, to drive fast wind streams \citep{suzuki2005making,shoda,shoda2019three, verdini2019turbulent}.  

Despite the vast body of work investigating parametric instabilities of parallel propagating (e.g., \cite{sakai, wong,nariyuki,Jayanti&Hollweg_1993}) and oblique propagating Alfvén waves \cite{vasquez1996formation,del2001parametricGRL,matteini2010parametric,DelZannaEA2015}, the heating processes resulting from the coupling between Alfv\'en and compressible modes remain poorly understood from a kinetic perspective. Among the most prominent issues still not fully investigated is how much of the energy carried by the pump wave  goes into internal energy, and how the latter is partitioned between protons and electrons. This is fundamental to understand the role of PDI in solar wind heating and acceleration. Indeed, in-situ and remote-sensing observations have found preferential heating of ions relative to the electrons \citep{wilson2018statistical, cranmer2020updated} and in general there is a well-known positive correlation between solar wind speed and proton temperature~\citep{elliott_2012}, whereas the solar wind speed is anti-correlated with the coronal electron temperature~\citep{geiss1995southern}. If PDI contributes significantly to coronal plasma heating, it  must be consistent with those constraints. %One of our goals is exactly to verify if energy partition during parametric decay is consistent with those observations.    

The goal of this work is  to investigate electron kinetic effects during the parametric decay of a parallel-propagating  Alfv\'en wave and to determine how the initial energy stored in the pump wave is partitioned between the internal energy of protons and electrons. So far, only a few works have studied parametric instabilities with a fully kinetic (PIC) approach \citep{sakai2005particle, nariyuki2008parametric}. Those works have considered reduced mass ratio ($m_i/m_e=16$) and a smaller wavelength for the pump wave. However, in order to understand if large scale processes can affect the electron population, one has to maintain a large scale separation between electrons and ions. A lower mass ratio would increase the relative electron gyroradius, possibly introducing artificial electron heating, e.g., via unrealistic particle scattering at magnetic structures. Here we perform PIC simulations of a weakly dispersive circularly polarized Alfv\'en wave with a more realistic electron-to-proton mass ratio in 1D; $m_i/m_e=400$ ensures that electron and ion scales are well separated while at the same significantly reducing the computational cost of resolving the electron scales (needed for the explicit PIC simulation) with respect with a realistic mass ratio simulation with the same size in terms of ion skin depths. We compare results from hybrid and particle-in-cell models, and consider different proton-to-electron temperature ratio and total plasma beta. For the PIC simulations we use the semi-implicit, energy conserving ECsim code \citep{lapenta2017exactly,lapenta2017multiple,gonzalez2018performance} and the explicit VPIC code \citep{bowers20080, bowers2008ultrahigh,bowers2009advances}. The semi-implicit algorithm  allows us to resolve only the scales of interest without resolving the Debye length as in explicit PIC codes. This feature will be useful in future work in 2D and 3D  when explicit simulations become prohibitively expensive. In addition, we run hybrid  simulations with the CAMELIA code \citep{matthews1994,Franci_2018} as a reference for the decay process and proton heating in the absence of electron kinetic effects. 

In section~\ref{model} we describe the model and simulation setup and in section~\ref{results} we present the main results. The conclusions of this study are summarized in section~\ref{discussion}.

\section{Model and simulation setup}
\label{model}

We consider a large amplitude, monochromatic Alfv\'en wave propagating parallel to the mean magnetic field ${\bf B}_0$, taken along the \textit{$x$}-axis. The initial perturbation for the electromagnetic field is given by $\delta B_y = \delta B \ \sin{(k_0 x)}$, $\delta B_z = -\delta B  \ \cos{(k_0 x)}$, $E_y = -\delta B \ \omega_0/(k_0 c) \ \cos{(k_0 x)}$ and $E_z = -\delta B \ \omega_0/(k_0 c) \ \sin{(k_0 x)}$. The current density is carried by both species, and is initialized in order to  satisfy Ampere’s law with an initial drifting Maxwellian distribution with mean proton and electron velocities given by
\citep{sakai2005particle}:

\begin{subequations}
\begin{equation}
\delta {\bf u}_i = \left( \frac{(\omega_0 + \Omega_{ce})(\omega_0^2 - c^2 k_0^2)}{(\omega_{pe}^2 + \omega_{pi}^2)k_0}  - \frac{\omega_0}{k_0}\right)  \frac{{\delta\bf B}}{B_0} 
\label{eq:New1} 
\end{equation}
\begin{equation}
\delta {\bf u}_e = \left( \frac{(\omega_0 - \Omega_{ci})(\omega_0^2 - c^2 k_0^2)}{(\omega_{pe}^2 + \omega_{pi}^2)k_0}  - \frac{\omega_0}{k_0}\right)  \frac{\delta{\bf B}}{B_0},  
\label{eq:New2}
\end{equation}
\label{Eq1}
\end{subequations}

where $\omega_{p\alpha}=\sqrt{e^2n_\alpha/\epsilon_0m_\alpha}$ and $\Omega_{c\alpha}=q_\alpha \mathbf{B}/m_\alpha$ are the plasma frequency and the gyro-frequency of the specie $\alpha=i,e$. The initial wave frequency $\omega_0$ is obtained from the dispersion relation of parallel propagating waves in a two-fluid plasma~\citep{baumjohann2012basic}:

\begin{equation}
\frac{k^2 c^2}{\omega^2} = 1 - \frac{\omega_{pe}^2}{\omega(\omega+\Omega_{ce})} - \frac{\omega_{pi}^2}{\omega(\omega-\Omega_{ci})}.
\label{eq:New3} 
\end{equation}

\begin{table}
  \begin{center}
\def~{\hphantom{0}}
  \begin{tabular}{cccccccccccccc}
    Run  & $L_x(d_i)$ & $\Delta x (d_i)$ & $\Delta t (\Omega_{ci}^{-1})$ &ppc&$\beta$&$\beta_i$&$\beta_e$&$T_i/T_e$\\[3pt]
    Hyb-TiTe1 & 102.4 & 0.1 & 0.005&10000&0.25&0.125&0.125&1  \\
    Vpic-TiTe1 & 102.4 & 0.0125&0.000125&500&0.25&0.125&0.125&1\\
    ECsim-TiTe1 & 102.4 & 0.025 & 0.0005& 500 &0.25&0.125&0.125&1 \\
    ECsim-TiTe1B & 102.4 & 0.025 & 0.0005 & 500 & 0.1575 & 0.078125 & 0.078125 & 1\\
    ECsim-TiTe4 & 102.4 & 0.025 & 0.0005 & 500 & 0.1575 & 0.03248&0.125&4
  \end{tabular}
  \caption{Initial conditions for the simulations presented in this paper.}
  \label{table1}
  \end{center}
\end{table}

In the rest of the paper we present our results using the following normalization: the magnetic field and density fluctuations are expressed in units of the guide field magnitude $B_0$ and of the initial plasma density $n_0 = n_{i_0} = n_{e_0}$, respectively.  Lengths are normalized to the proton inertial length $d_i =  c/\omega_{p_i}$, time is expressed in units of the inverse of proton gyrofrequency $\Omega_{ci}^{-1}$ and the velocities are normalized to the proton Alfv\'en speed $v_A = B_0/\sqrt{4 \pi n_0 m_i}$. The plasma beta for both protons and electrons is defined as $\beta_{i,e}= 2 n_0 k_B T_{i,e}/B_0^2$. The electron characteristic scales are related to the proton scales through the mass ratio that we fixed to $m_i/m_e=400$ for the PIC simulations. We considered an initial weakly dispersive pump wave with a wave number $m_0=4$ and wave vector $k_0 = 2\pi m_0/L$,  $L$ being the size of the simulation domain. The initial wave amplitude is $\delta B/B_0=0.5$ and the wave propagates with a phase velocity $\omega_0/k_0 = 0.885 v_A$, with $v_A/c=0.05$ and $k_0 d_i = 0.2454$. A similar setup for the magnetic and the velocity fields is used to initialize the hybrid simulation (in the limit $\Omega_{ce}\gg \omega$). A summary of the numerical parameters for the simulations presented here can be found in Table~\ref{table1}. We compare the results of simulations with the same plasma beta and different temperature ratio between electrons and protons (ECsim-TiTe1B and ECsim-TiTe4) with simulations  with same temperature ratio and different beta (ECsim-TiTe1 and ECsim-TiTe1B), to separate the effects of the two parameters. 

\section{Results}
\label{results}
\subsection{Overview}

\begin{figure*}
\includegraphics[width=0.9\textwidth]{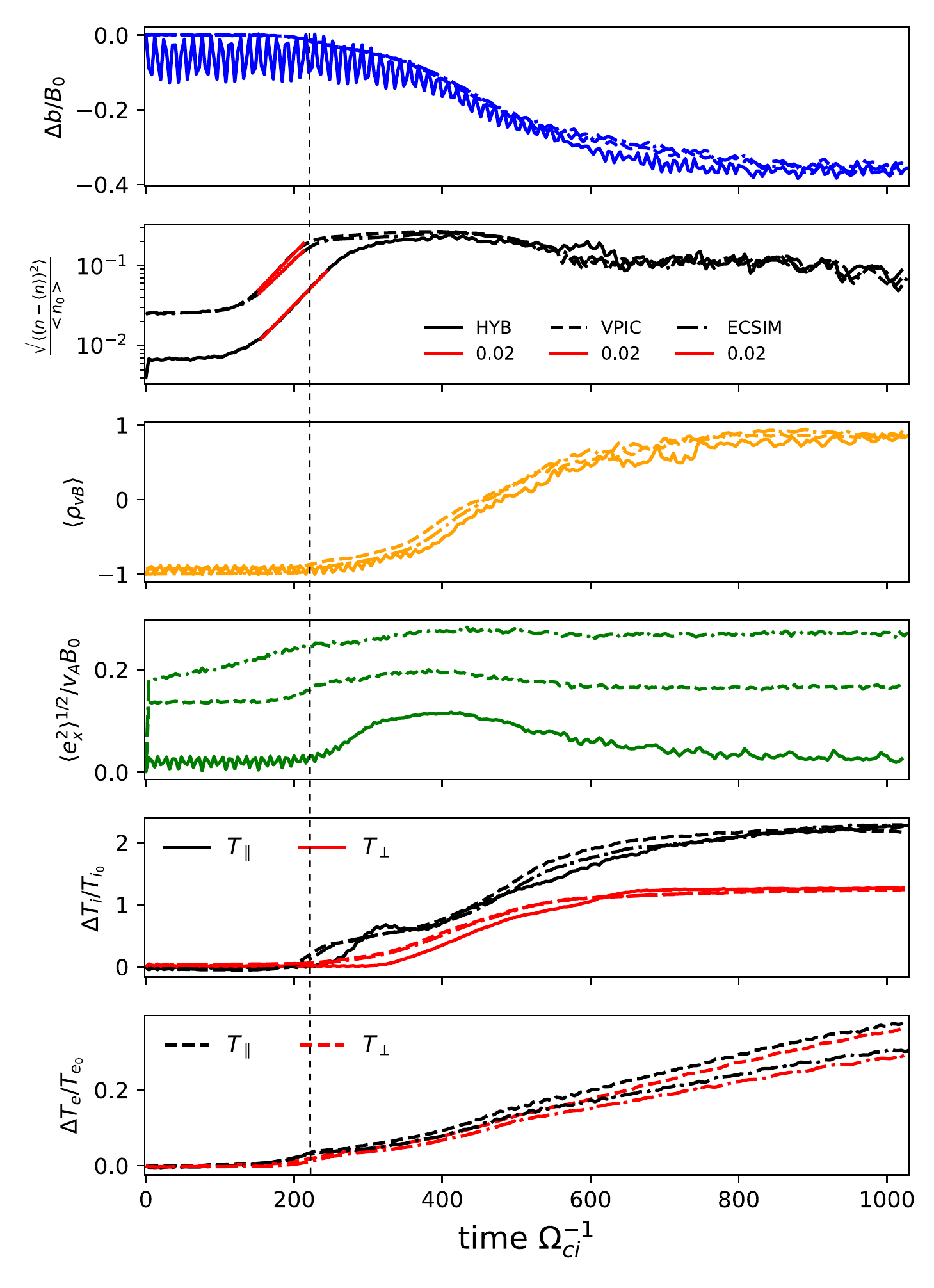}
\caption{Temporal evolution of the variation of the rms of magnetic field fluctuations (top panels), rms of total density fluctuations (second panels), the correlation between magnetic and velocity fluctuations $\rho_{VB}$ (third panel), and rms of the field-aligned electric field (fourth panel). The last two panels show the variation of the proton and electron mean parallel and perpendicular temperatures. The left column shows a comparison between runs Hyb-TiTe1, Vpic-TiTe1, and Ecsim-TiTe1. The vertical black dashed lines represent the end of the linear stage of PDI for Ecsim-TiTe1.}
\label{Fig1a}
\end{figure*}

\begin{figure*}
\includegraphics[width=0.9\textwidth]{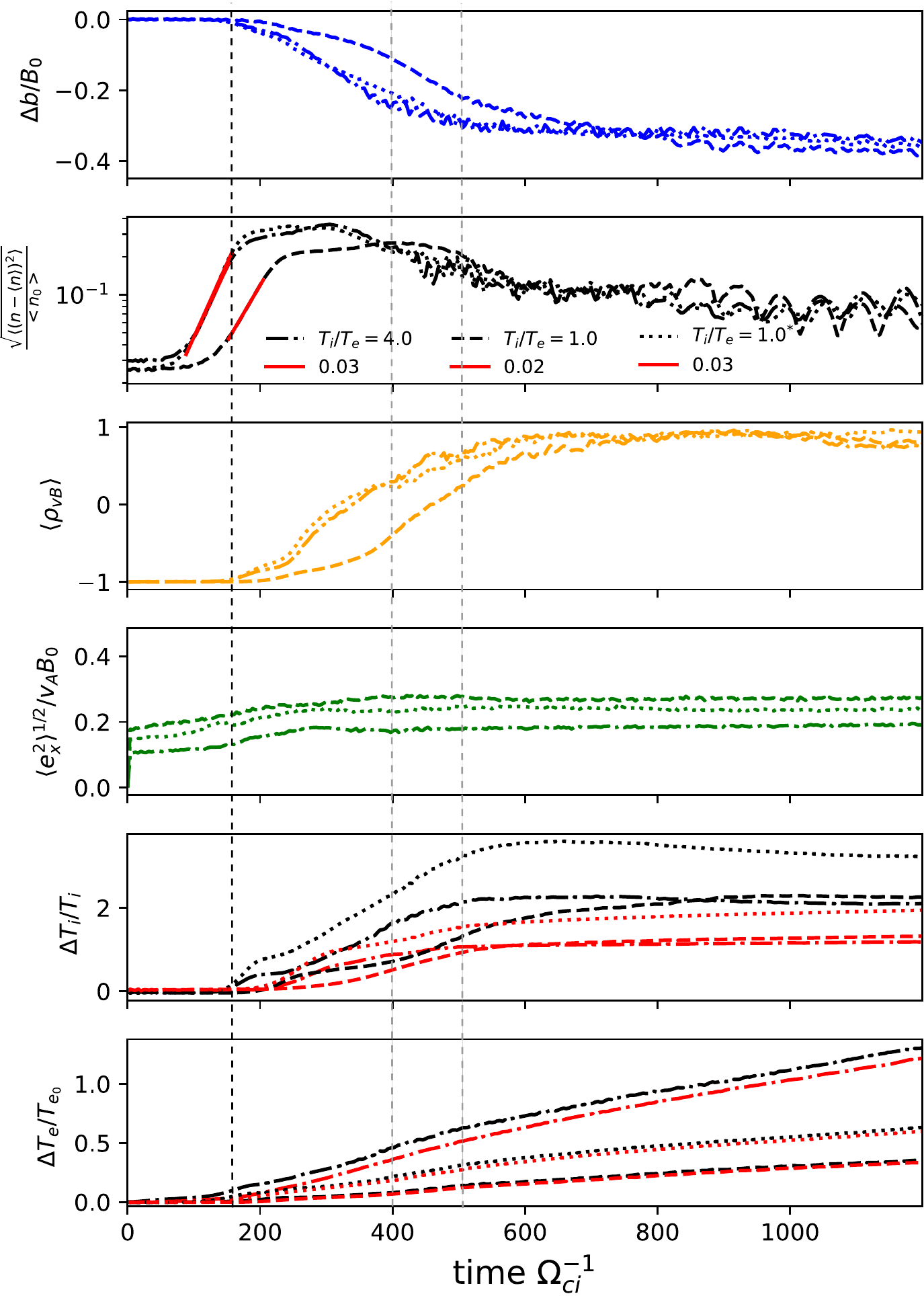}
\caption{Temporal evolution of the variation of the rms of magnetic field fluctuations (top panels), rms of total density fluctuations (second panels), the correlation between magnetic and velocity fluctuations $\rho_{VB}$ (third panel), and rms of the field-aligned electric field (fourth panel). The last two panels show the variation of the proton and electron mean parallel and perpendicular temperatures. Results for runs Ecsim-TiTe4 (dot-dashed line), Ecsim-TiTe1 (dashed line) and Ecsim-TiTe1B (dotted line). The vertical black dashed lines represent the end of the linear stage of PDI for Ecsim-TiTe4 and Ecsim-TiTe1B. The gray dashed lines mark the times when distribution functions are shown in Fig.~\ref{Fig3}.}
\label{Fig1b}
\end{figure*}

Figure~\ref{Fig1a} and 
Figure~\ref{Fig1b} provide an overview of the time evolution of the parametric decay in the simulations listed in Table~\ref{table1}. We show, from  top to  bottom,  the variation of the normalized root mean square (rms) of the magnetic and total density fluctuations, the correlation between velocity and magnetic field fluctuations ($\rho_{vB} = \mathbf{v} \cdot \mathbf{B} / \lVert \mathbf{v} \rVert \lVert \mathbf{B} \rVert$), the rms of the field aligned electric field $e_x$,  and the variation of the mean  proton and electron temperatures  normalized to their initial value. The growth rate of the PDI is reported in the legend and corresponds to the slope of the best  fit indicated by the red solid line in the second panels. 

In Fig.~\ref{Fig1a} we compare results from the hybrid simulation and from the explicit (VPIC) and the semi-implicit (ECsim) PIC simulations, for $T_i=T_e$ and $\beta_e=0.125$. The decay of the pump wave, marked by the exponential growth of density fluctuations (second panel), proceeds similarly in the hybrid and PIC simulations. In both models the decay starts at the same time, around $t\Omega_{ci}\simeq 150$, with the density fluctuations growing at the same rate.  The saturation of the instability occurs when the backward propagating Alfvén wave and forward propagating ion-acoustic wave are well developed and density fluctuations reach a constant value of $\delta n_{rms}/<n>\simeq 0.2$ at around $t\simeq 225 \Omega_{ci}^{-1}$ (vertical dashed line). A complete reflection of the pump wave then takes place, as can be seen from the third panel, when the sign of $\rho_{VB}$ changes from $-1$ to $+1$. During this stage, a portion of the initial magnetic energy is converted into kinetic and internal energy of both protons and electrons. % PIC simulations show that  most of the heating goes into protons (fifth and bottom panels). 
In particular, protons undergo strong parallel heating along the local magnetic field with significant perpendicular heating (fifth and bottom panels), as reported in previous studies \citep{Gonzalez_2020}. PIC simulations show that electrons are not isothermal during the decay process, and that they are also energized. The heating of electrons is on average isotropic and proceeds at nearly a constant rate. This can be seen from the bottom panels of Fig.~\ref{Fig1a}, showing that $T_{e\parallel}\simeq T_{e\perp}$ and that temperatures keep increasing during the non-linear stage (specifically after $t\Omega_{ci}\simeq 500$). We noticed a small difference in the electron temperature for the explicit and semi-implicit simulations with a slightly larger final temperature in the VPIC simulation. This might be due to numerical heating associated with explicit codes, although the relative error of the total energy is less than $1\%$ in both simulations.  

The Fig.~\ref{Fig1b} shows results for simulations performed with ECsim with different $\beta$ and proton to electron temperature ratio $T_i/T_e$.
As can be seen from these plots, the total plasma $\beta$ determines the growth rate and the overall decay of the pump wave in agreement with fluid theory \citep{wong1986parametric,hollweg1994beat}, which predicts that higher beta plasmas have a lower growth rate. This dependence can be seen by comparing runs ECSim-TiTe4 (dot-dashed) and TiTe1B (dotted), which have different electron temperature but same total plasma beta, and by comparing ECSim-TiTe1B and TiTe1, corresponding to different values of the plasma beta.  We observe that the higher the initial $T_e$, the higher the electric field, in agreement with the  dependence of the electric field on $\beta_e$ from the generalized Ohm's law.

The most significant difference between the simulations with same beta but different $T_i/T_e$ is registered in the rms of the electric field, (second column, fourth panel: dotted  vs. dot-dashed line). We observe lower rms values of  $e_x$ when $T_i>T_e$. This is consistent with Landau damping of the ion-acoustic mode, which is expected to be stronger when $T_i>T_e$ than in the case where $T_i=T_e$. Linear Landau damping however is not strong enough to affect the growth of the instability. 

\begin{figure}
\includegraphics[width=0.95\textwidth]{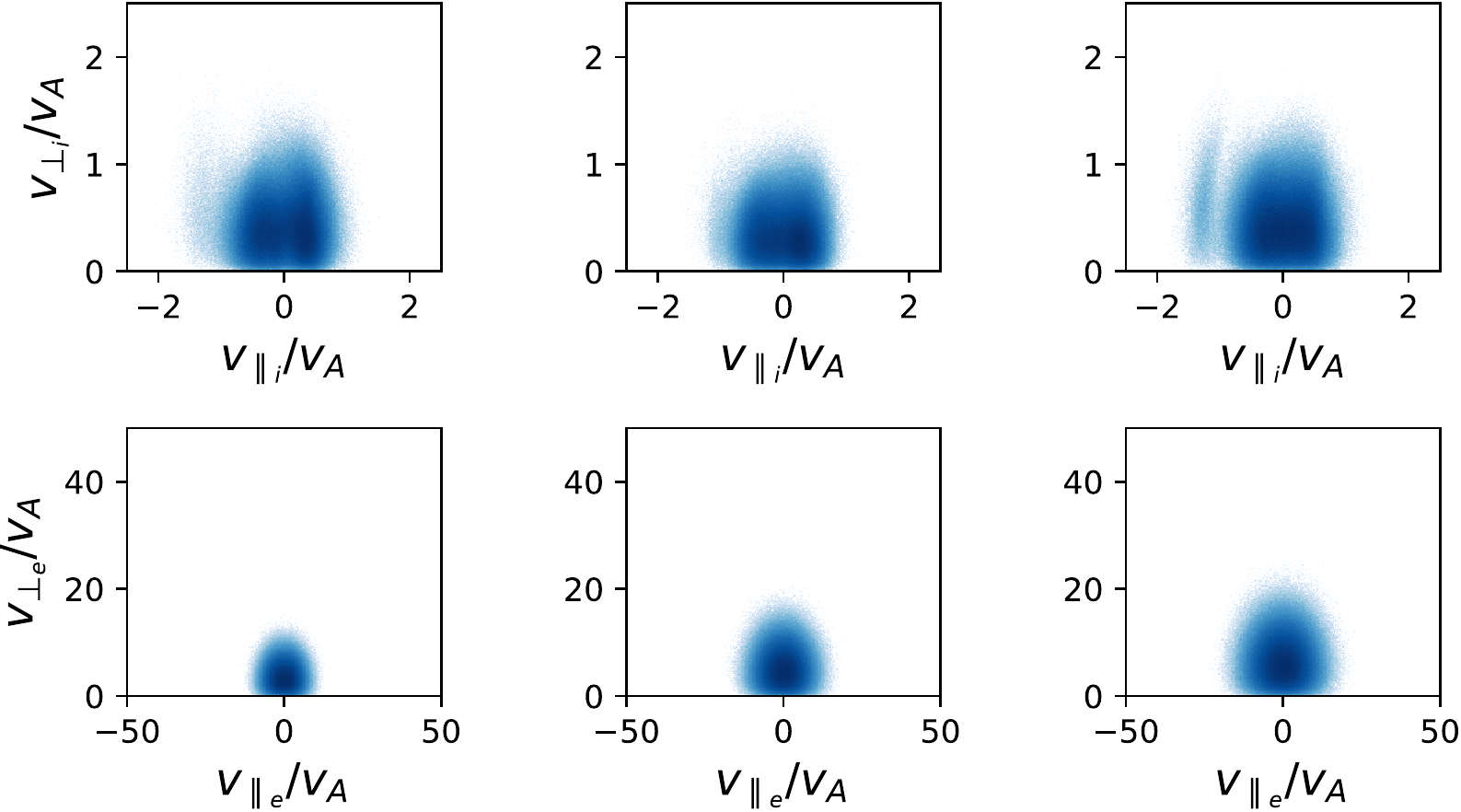}
\caption{Reduced particle VDF for protons (top) and electrons (bottom) integrated over the entire domain. We show, from left to right, the VDFs for  Ecsim-TiTe4, Ecsim-TiTe1 and Ecsim-TiTe1B, at $t=400 \Omega_{ci}^{-1}$, $t=560 \Omega_{ci}^{-1}$ and $t=400 \Omega_{ci}^{-1}$, respectively.}
\label{Fig3}
\end{figure}

To conclude our overview of the decay process, in Fig ~\ref{Fig3} we show the contour plots of the velocity distribution functions (VDF) in velocity space, $f(v_\parallel,|v_\perp|)$, integrated in $x$. We plot the VDF of protons (top panels) and electrons (bottom panels) when the system has  achieved the nonlinear stage for runs ECSim-TiTe4, ECSim-TiTe1 and ECSim-TiTe1B. This time  is marked by the vertical gray dashed lines in the right panel of Fig.~\ref{Fig1b}, at $t=400 \Omega_{ci}^{-1}$, $t=560 \Omega_{ci}^{-1}$ and $t=400 \Omega_{ci}^{-1}$ for each simulation, respectively. As will be discussed in the next sections, protons develop an anisotropic distribution function with a clear field-aligned beam propagating backward, and an energized field-aligned population at the (positive) ion-acoustic speed (see Fig.~\ref{Fig4} for more details). Electrons instead remain on average isotropic, and their heating is much less pronounced than proton heating.

\subsection{Energy conversion and particle heating}

\begin{figure*}
\includegraphics[width=\textwidth]{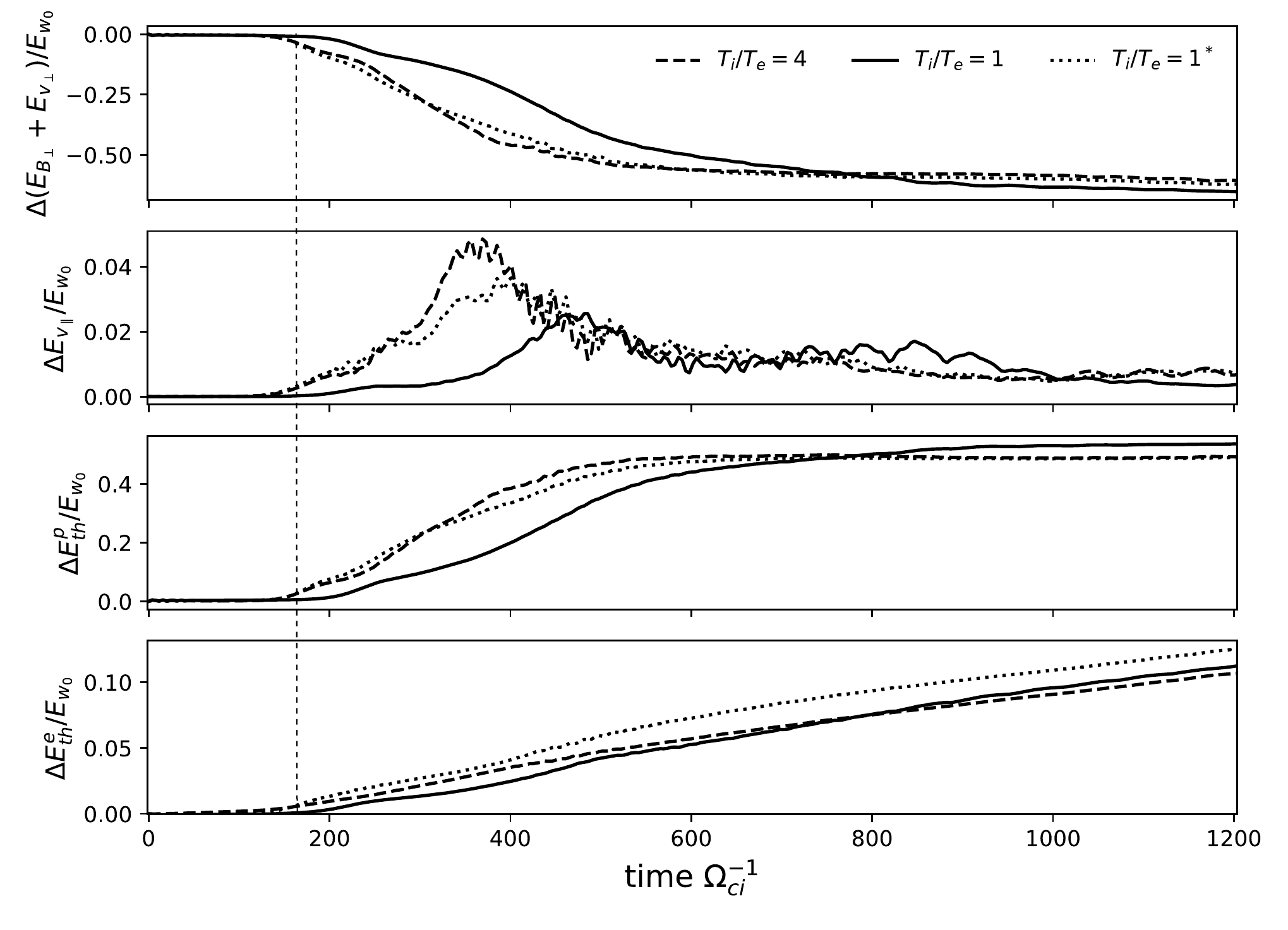}
\caption{Energy conversion and partition.  Variation of the wave energy (top panel),  of the field-aligned bulk kinetic energy (second panel), and  the internal energy variations for protons and electrons (third and fourth panels, respectively). Each panel shows results for runs Ecsim-TiTe4 (dashed), Ecsim-TiTe1 (solid) and Ecsim-TiTe1B (dotted). The vertical dashed line indicates the end of the liner stage for the case $T_i/T_e=4$.}
\label{Fig2n}
\end{figure*}

The energy lost by the pump wave powers the formation of the daughter waves and results into thermal and parallel bulk energy gain for protons and electrons. PIC simulations show that, after the decay of the pump wave, most of the particle internal energy gain goes into proton internal energy. In Fig.~\ref{Fig2n}, we analyze the pump wave energy conversion and partition by breaking down the total energy (per unit volume) into its contributions from Alfv\'en wave energy (top panel), parallel kinetic energy (thus including the contribution from the ion acoustic wave; second panel) and internal energy of protons and electrons (third and bottom panels). All quantities are expressed as variations from their initial value, and they are normalized to the initial pump wave energy $E_{w0}=E_{B_\bot}(0)+E_{v_\bot}(0)$, where %
$$E_{B_\bot}(t)=1/2<B_y(t)^2+B_z(t)^2>$$ 

is the magnetic energy density and 
$$E_{v_\bot}(t)= \sum_\alpha 1/2 < m_\alpha n_\alpha (u_{y_\alpha}(t)^2+u_{z_\alpha}(t)^2) >$$ 

is the bulk kinetic energy and the brackets denote spatial average. The $x$-component of the kinetic energy is defined as 
$$E_{v_\parallel}(t)= \sum_\alpha 1/2< m_\alpha n_\alpha u_{x_\alpha}(t)^2>,$$ 

and the internal energy of species $\alpha$ as 
$$E_{th,\alpha}=1/2< Tr(P_{ij,\alpha})>.$$ 

Energy balance is quantitatively similar in all our simulations. The top panel of Fig.~\ref{Fig2n} shows that the energy contained in the Alfv\'en waves decreases of  $|\Delta E_w|/E_{w0} \simeq 60\%$. Notice that this does not correspond to the total pump wave energy decrease. The pump wave fully decays in the simulations considered here (see the correlation $\rho_{vb}$ in Fig.~\ref{Fig1b}). After the decay ($\rho_{vb}=+1$ after $t\Omega_{ci}\gtrsim 500$), $40\%$ of the pump wave energy goes into a reflected Alfv\'en wave, and the remaining $60\%$ goes in an ion-acoustic mode and other forms of internal energy. The second panel, as well as the trend of $\rho_{rms}$ reported in Fig.~\ref{Fig1b}, shows that at the end of the linear stage (indicated by the vertical dashed line for the case $T_i/T_e=4$), $\Delta E_\parallel$ has increased due to the growth of the ion-acoustic mode. The subsequent peak of  $E_\parallel$ however is due to the ongoing particle trapping by the ion-acoustic mode that generates a population of protons moving at about the ion-sound speed, as discussed in section~\ref{phase_space_dyn} and, e.g., in ~\cite{MatteiniEA2010}. Particle trapping and subsequent phase space mixing ultimately lead to  the dissipation of the ion-acoustic mode and particle heating. After the complete decay of the pump wave, most of the Alfv\'en wave energy has  gone into proton internal energy. In particular, we find that about $50\%$ of the initial pump wave energy goes into proton internal energy and $10\%$ to electron internal energy. Notice that, unlike protons, electrons keep heating at a constant rate the instability has saturated and the pump wave has decayed. At the end of the simulation, the electron internal energy is a factor of 2 larger than the energy at the end of the complete decay ($t\Omega_{ci}\gtrsim 500$). Still, the good energy conservation in the simulations and similar results between explicit and semi-implicit simulations suggest that the electron heating is indeed physical and not a numerical artifact. Such an energy balance appears to be unaffected by the total plasma beta and only slightly by $T_i/T_e$ and $m_i/m_e$ (not shown). 

%Contour plot of different quantities in the simulations
\begin{figure*}
\includegraphics[width=\textwidth]{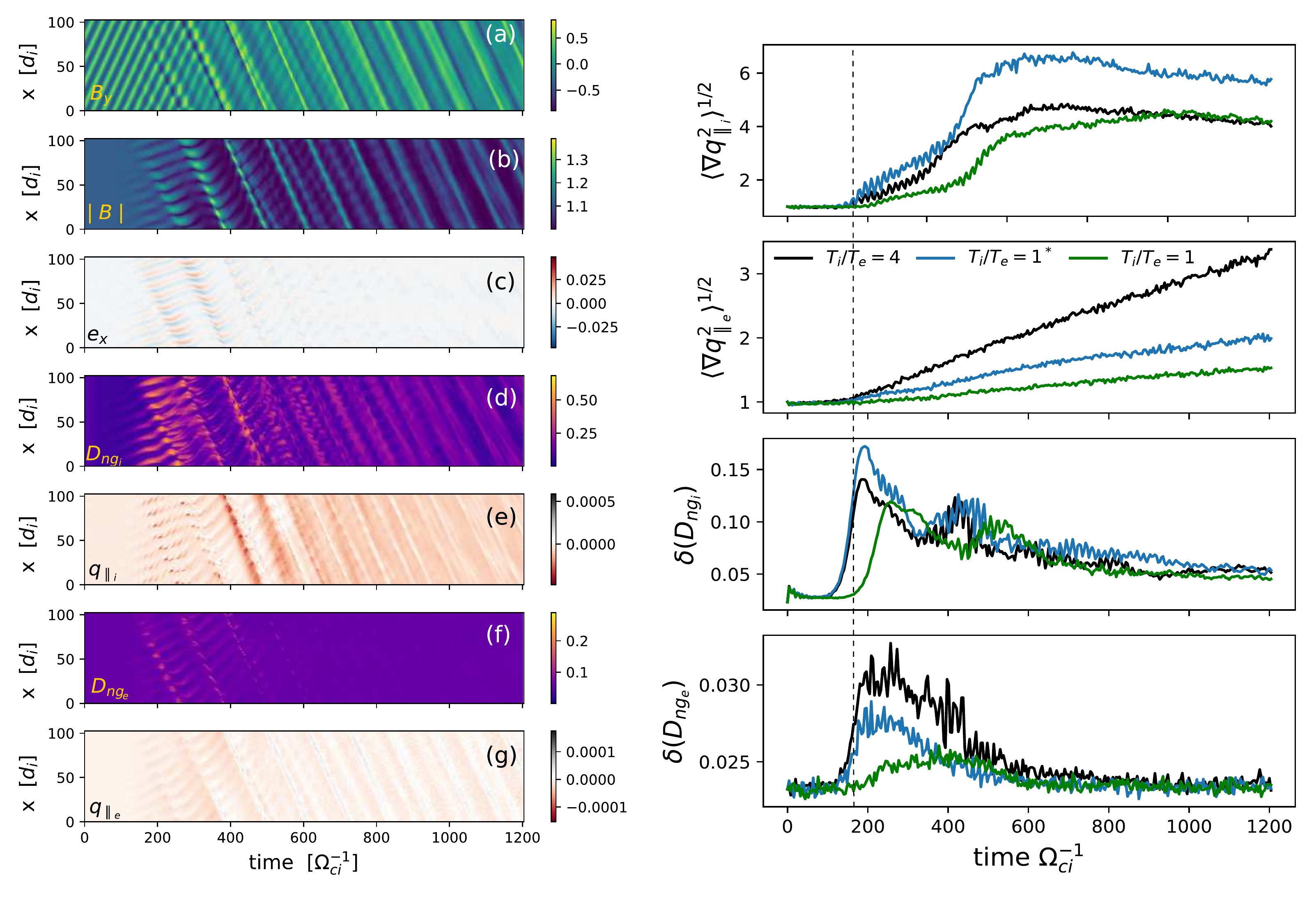}
\caption{Left panels:  contour plot in the $(t,x)$ plane of the $b_y$ component (panel (a)), the magnitude of the magnetic field (panel (b)), the field-aligned component of electric field (c), and the scalar agyrotropy and heat flux for protons and electrons (panel (d) and (e)) and (panel (f) and (g)), respectively. Results are shown for ECSim-TiTe4. Right panels: rms of the divergence of the parallel heat flux for protons (top panel) and electrons (second panel), and variance of the agyrotropy for protons (fourth panel) and electrons (bottom panel) for all simulations.}
\label{Fig2}
\end{figure*}

\subsection{Phase space dynamics}
\label{phase_space_dyn}

Compressibility plays an important role in proton and electron energization and in providing  paths to particle heating. In the low-beta simulations presented in this work, we were able to identify and distinguish two types of compressible fluctuations generated during parametric decay: an ion-acoustic mode and a compressible steepened front of the fast-mode type propagating at the Alfv\'en speed. Such signatures were difficult to discern in higher beta simulations \citep{Gonzalez_2020}. %As we are going to discuss in this section, 
Both types of compressible fluctuations contribute to the increase of internal energy. However, they lead to fundamentally different phase space dynamics and to different kinetic features. The ion-acoustic mode contributes nonlinearly to phase space mixing via particle trapping. On the other hand, the development of the steepened front with non-constant $|{\bf B}|$ leads to the formation of strong parallel heat fluxes and ultimately causes the acceleration of a population of protons into a well-defined beam of particles and the enhancement of proton perpendicular heating. In the following, we analyse the signatures of these processes in the fields and particle VDFs. 

In Fig.~\ref{Fig2}, left column, we show contour plots in the $(t,x)$ plane of $B_y$  (top panel), of the magnetic field magnitude $|{\bf B}|$ (second panel) and of the  electric field $e_x$ (third panel). We also plot the scalar pressure agyrotropy \citep{aunai2013electron, cazzola},
$$D_{ng}=\sqrt{8\left( P_{xy}^2 + P_{xz}^2 + P_{yz}^2 \right)}/(P_\parallel + 2 P_\perp),$$

and the parallel heat flux ($q_\parallel$) for protons (fourth and fifth panels) and  electrons (sixth and bottom panels), respectively. The thermal energy flux is obtained by calculating the total energy flux for each specie ($\mathbf{Q}_s = \frac{1}{2}m_s \int f_s(\mathbf{x},\mathbf{v},t) v^2 \mathbf{v} d^3\mathbf{v}$) and subtracting the energy flux components in the Eulerian frame ($\mathbf{q}_s = \mathbf{Q}_s -   \mathbf{H}_s - \mathbf{K}_s$), namely  the contribution from the bulk energy flux ($\mathbf{K}_s = \mathbf{u}_s \frac{\rho u_s^2}{2}$) and the enthalpy flux ($\mathbf{H}_s = P_s \cdot \mathbf{u}_s + \mathbf{u}_s U_{th}$). This requires the calculation of the lower moments of the velocity distribution (see details in \cite{lapenta2020multiscale}). We show results from ECSim-TiTe4 as a reference, since ECSim-TiTe1 and ECSim-TiTe1B are similar. In the right column of Fig.~\ref{Fig2} we show the rms of the divergence of the parallel heat flux (normalized to its initial value) and the standard deviation of the scalar agyrotropy for both species and for all simulations.

From Fig.~\ref{Fig2}, left, one can see that a dominant backward propagating mode forms at $t\Omega_{ci}\simeq 200$, when the PDI has reached saturation. The signature of the forwards ion acoustic mode can be seen in the plots of $e_x$ in the time interval $t\Omega_{ci}=200-400$, before it dissipates, corresponding to a wavevector $k_s=7 \times 2\pi/L_x$. At about $t\Omega_{ci}\simeq 400$, a steepened front propagating backwards at the Alfv\'en speed also develops. This steepened front corresponds to a localized increase of $|\mathbf{B}|$, positively correlated with a density enhancement. As discussed in our previous work \citep{Gonzalez_2020,Gonzalez_2021},  compressive fluctuations in  $|{\bf B}|$ are associated to a field-aligned electric field $e_x$. Enhancements of $e_x$ at the steepened front can also be seen in the third panel. 
The strongest non-thermal features (agyrotropy and heat fluxes) of particles develop at such a steepened front and within the ion-acoustic mode, as shown in Fig.~\ref{Fig2},  from the third to the bottom panels on the left.  

The cumulative contribution of non-Larmor particle motion produces non-zero values of the off-diagonal terms of the pressure tensor, quantified by $D_{ng}$, where local values of up to  $|D_{ng,e}|\simeq 0.3$ and $|D_{ng,i}|\simeq 0.65$ are reached. The variance of $D_{ng}$ for both species is reported in the third and fourth panels on the right, and the rms of the divergence of the parallel heat flux in the first two panels. Agyrotropy in particular is considered one of the VDF based dissipation measures \citep{pezzi2021dissipation}. From Fig.~\ref{Fig2}, right panels, one can see that agyrotropy first  builds up during the linear stage of the PDI for both species in correspondence with the ion-acoustic mode, before decaying during the nonlinear stage as the ion-acoustic mode is dissipated and, correspondingly, the plasma is heated. However, protons, unlike electrons,  undergo a two-step heating process. Besides particle trapping, protons subsequently interact with the steepened front, where they develop a second peak of agyrotropy. Interaction of protons  with a field-aligned discontinuity and electric field causes both proton scattering in phase space, contributing to heating, and the acceleration of protons into a field-aligned beam \citep{Gonzalez_2021}. This is consistent with the observed time evolution of proton agyrotropy and parallel heat flux (see right top and third panels in  Fig.~\ref{Fig2}).  Electrons, which have a smaller gyroradius than protons, display on average  $D_{ng,e}\ll D_{ng,i}$ (bottom right panel) and, accordingly, they do not gain much internal energy.  

\begin{figure}
\includegraphics[width=\textwidth]{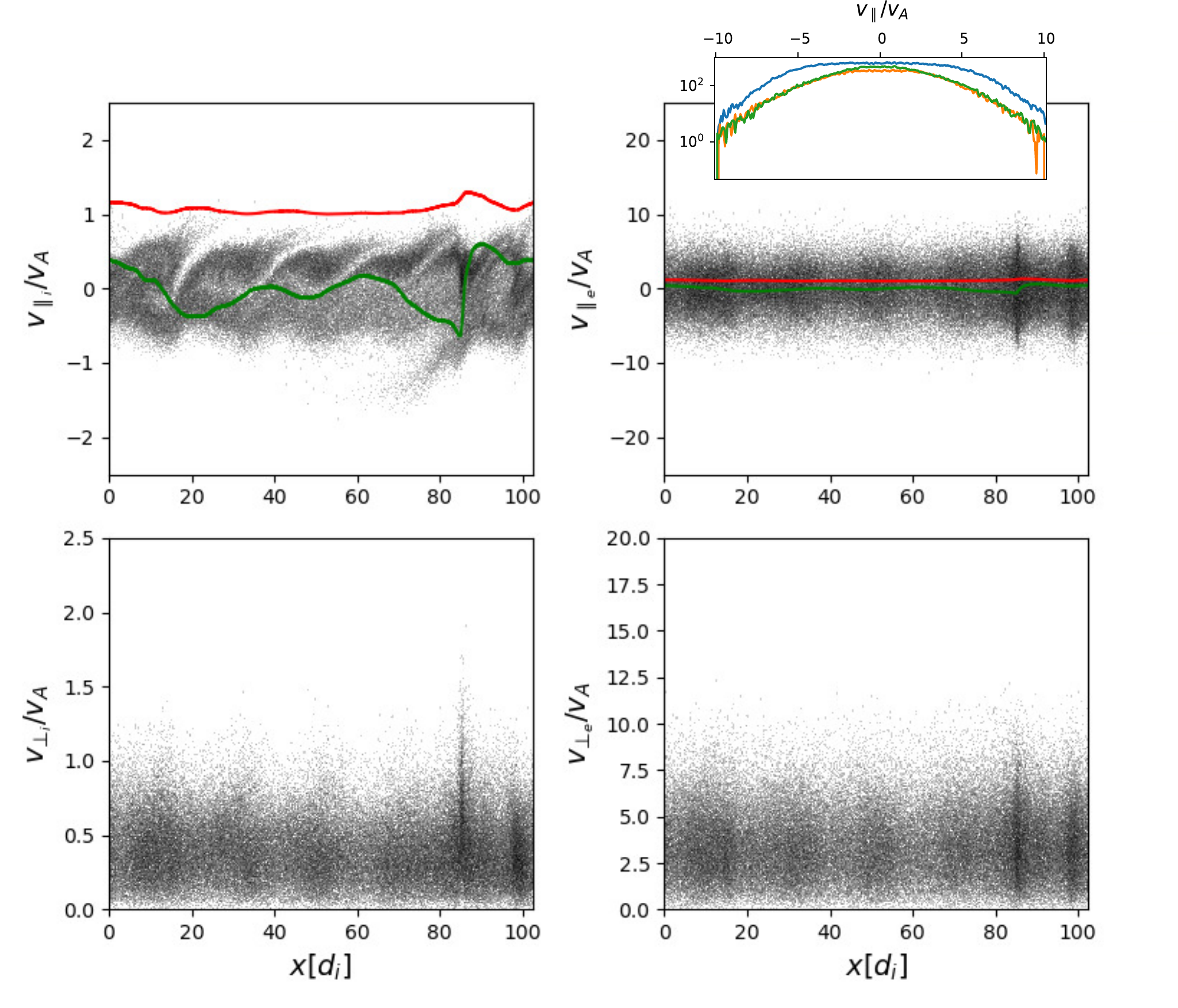}
\caption{The phase-space ($x,v_\parallel/v_\bot$) for protons (left) and electrons (right). The phase space is shown at $t\Omega_{ci}=400$ for Ecsim-TiTe4. The magnetic field’s magnitude  (red line) and  $B_y$ (green line) are also presented. The inset in the top right panel represents the electron distribution function integrated in space around the discontinuity (blue) and outside the discontinuity (orange and green).}
\label{Fig4}
\end{figure}

Kinetic features in  phase space corresponding to the dynamical evolution described above can be seen in Fig. \ref{Fig4}, where we show the proton and electron VDFs in the $(x,v_\parallel)$ and $(x,v_\perp)$ space, top and bottom panels, respectively, at $t\Omega_{ci}=400$. We show results for EcsimTiTe4 only since similar features develop in the other simulations. Protons in phase space are shown on the left, and electrons on the right.   For reference, we  plot  $B_y(x)$ (green color) and $|\mathbf{B}(x)|$ (red color) at the same time to show that the particle beam forms at the steepened front. The contour plot of  $f(x,v_\parallel)$ shows the two sources of the proton internal energy increase discussed above. First, saturation of the growth of the ion-acoustic mode is  determined by nonlinear particle trapping. The latter causes the formation of a second proton population that propagates forwards at the ion-acoustic speed, and develops  well-known signatures of phase space mixing (vortexes). Second, a proton beam is generated locally at the backward propagating steepened Alfv\'en front. The latter can be identified in Fig.~\ref{Fig4} at around $x \simeq 85 \ d_i$, and has a size of a few $d_i$.  Both particle trapping and beam formation cause the large increase of the parallel proton temperature and parallel heat flux. One can see in the bottom left panel that also perpendicular heating takes place at the steepened edge, where particles with high perpendicular velocities are observed. The perpendicular energization is due to particle scattering at those discontinuities \citep{Gonzalez_2021}. The electrons are also characterized by a local deviation from Maxwellian at the steepened front. The inset in the top right panel shows the electron VDF integrated in $x$ over an interval $\Delta x=3 \ d_i$, at different positions in the simulation domain: centered at the location of discontinuity $x=85 \ d_i$ (blue line) and outside the discontinuity, at $x=60 \ d_i $ and $x=95 \ d_i$ (orange and green lines), where the VDF is Maxwellian. The electrons at the discontinuity show a flat-top distribution with enhanced number of electrons in that region as  to maintain the quasi-neutrality of the system (we do not find evidence of charge-space separation in the simulations). We do not observe any electron beam in the simulations, and the electrons undergo on average isotropic heating.  

\section{Conclusions}
\label{discussion}

We have investigated the parametric decay instability of a  large-amplitude Alfvén wave with large-scale particle-in-cell simulations with an explicit (VPIC) and semi-implicit (ECsim) PIC code. Results from the PIC model are compared with those from a state-of-the-art hybrid model (CAMELIA) that employs an isothermal electron closure. Our main findings can be summarized as follows:

\begin{itemize}
    \item  The growth rate of the decay is determined by the total plasma beta in agreement with fluid theory. The electron temperature affects compressibility in a way that is consistent with the generalized Ohm's law. The higher $T_e$, the larger the rms of the field aligned electric field, and the lower the rms of density fluctuations. 
    \item Compressible effects lead not only to the generation of the ion acoustic mode, but also to the development of a fast-mode steepened front propagating with the backward Alfv\'en wave. Particle interaction with those two types of compressible fluctuations leads to the increase of particle internal energy.
    \item Proton internal energy gain has contributions from: $(i)$ phase space mixing due to particle trapping by the ion-acoustic wave; $(ii)$ beam acceleration aligned to the mean magnetic field at the steepened front;  $(iii)$ particle scattering at the steepened front. Protons are overall heated preferentially in the parallel direction due to the contributions of particle trapping and beam formation. Electrons are also energized at the steepened front. Contrary to protons, electrons are heated isotropically.    
    \item Analysis of the energy balance  shows that after the complete decay of the pump wave $40\%$ of the pump wave goes into reflected wave, $\sim50\%$ into proton internal energy (perpendicular and parallel)  and $\sim10\%$  into electron internal energy.  
\end{itemize}

In conclusion, we have clarified that parametric decay can heat the plasma via two  processes that in prior work \citep{araneda2008proton, MatteiniEA2010, Gonzalez_2020} were not as clearly separated as in this set of simulations. The processes are proton trapping and beam acceleration at steepened fronts. While proton trapping causes the energization of a proton population which propagates at the ion-sound speed, it is the steepened front that generates an actual beam at the Alfv\'en speed. Since the steepened front propagates backward due to the complete reflection of the pump wave, the beam also propagates backward in this set of simulations. Our results also provide the first energy balance analysis of parametric decay with fully kinetic simulations. In particular, we find that parametric decay preferentially heats protons, although electrons also gain a fraction of the  energy stored in the pump wave. Heating triggered by parametric decay is consistent with the observed preferential proton heating with respect to electrons in the corona. In this context, our results provide a promising starting point to assess the role of (kinetic) compressible effects in wave-driven theories of the solar wind.      

\acknowledgements{This research was supported by NASA grant \#80NSS\-C18K1211.  We also acknowledge NSF award 2141564 and the Texas Advanced Computing Center (TACC) at The University of Texas at Austin for providing HPC resources that have contributed to the research results reported within this paper. URL: http://www.tacc.utexas.edu. We thank G. Lapenta and the ECSim development team for granting us access to the code and .  M.E.I. acknowledges support from the German Science Foundation DFG within the Collaborative Research Center SFB1491.}

\bibliographystyle{jpp}
\bibliography{jpp-instructions}

\end{document}